\documentclass[doublecol,floats]{epl2}
\usepackage{amssymb}
\usepackage{graphicx}
\usepackage{dcolumn}
\usepackage{amsmath}
\usepackage{bm}
\usepackage{epsfig}

\newcommand{\be}{\begin{equation}}
\newcommand{\ee}{\end{equation}}

\def\NN{\mathbb{N}}
\def\PN{\mathbb{E}}
\def\QN{\mathbb{O}}

\title{Emergence of chaos in interacting communities}
\author{M. Ostilli \inst{1,2}, W. Figueiredo \inst{1}}
\institute{ \inst{1}
Departamento de Fisica, Universidade Federal de Santa Catarina, Florianopolis, 88040-900, SC, Brazil \\
\inst{2}
Dipartimento di Fisica, Universit\`a di Roma ‘La Sapienza’, Piazzale Aldo Moro 2, Roma I-00185, Italy
} 


\abstract{
We introduce a simple dynamical model of two interacting communities whose elements
are subject to stochastic discrete-time updates governed by only bilinear interactions.
When the intra- and inter-couplings are cooperative,
the two communities reach asymptotically an equilibrium state.
However, when the intra- or inter-couplings are anti-cooperative, 
the system may remain in perpetual oscillations and,
when the coupling values belong to certain intervals, 
two possible scenarios arise, characterized either by erratic aperiodic trajectories and high sensitiveness to small changes of the couplings,
or by chaotic trajectories and bifurcation cascades.
Quite interestingly, we find out that even a moderate consensus in one single community
can remove the chaos. 
Connections of the model with interacting stock markets are discussed.
}

\begin{document}

\maketitle


\textit{Introduction and Model Formulation.} 
Chaos is a fundamental paradigm and  
its analysis may have remarkable practical impacts \cite{StrogatzC,Lorenz,Baker}. 
For example, certain stock market models that exhibit chaos
succeed in explaining ``high volatility'' and ``bubbles'' \cite{Volatility}. 
More recently, it has been argued that, in some chaotic systems,
identifying specific signals might be useful to forecast and even prevent extreme events \cite{Dai,Gauthier}.    
In general, given a dynamical system described by a set of differential equations, a necessary condition for the onset of chaos 
is the existence of non-linear terms which, when referred to models governed by a potential energy only 
(\textit{i.e.} in the absence of a mechanical kinetic term), 
means that at least three-body interactions, possibly also self-body-like, are present.
An important example is the Kuramoto model \cite{KuramotoC,Strogatz,Pikovsky} 
which involves sinusoidal interactions and, as a consequence, a chaotic behavior
may set in for intermediate values of the couplings.
Chaos scenarios may also take place when the system is coupled to external random fields as in \cite{KuramotoR,Volatility}. 

In recent years, within complex networks theory, it has been recognized that most of real world networks are characterized
by the presence of a community structure that plays a major role in the functionality of the network \cite{Santo}.
In fact, many non trivial phenomena running on complex networks, \textit{e.g.} mutual antiferromagnetism, non standard percolation with
metastable states, specific spreading of diseases, etc..., are due to the presence of at least two communities \cite{Suchecki,Comm,Suchecki2,Stanley,Zhang}~
\footnote{More recently, it has been emphasized that networks of networks should be considered in order to take into account
the presence of links of different nature and role \cite{Multilayer}.}.
In this Letter, we analyze how the presence of two communities affects the dynamics generated by a minimal model.
Specifically, we show that, in a discrete-time dynamics
governed by only two-body interactions, 
the presence of two interacting communities  
supplies for the non-linearities or the external fields that lead to chaos. 
We stress that our model formulation is a microscopic one. The mean-field limit case allows
then to reduce the involved equations to a macroscopic dynamics from which an expertise
of chaos might recognize the emergence of chaos (see Eqs. (\ref{mastm1}) and (\ref{mastm2})).
Notice, however, that the microscopic model formulation, its connection with a real world phenomena,
and the role played by the two communities, responsible for a chaotic dynamics, are \textit{a priori} non obvious
and deserve a special attention.

Let be given two communities \cite{Santo} of agents $\mathcal{N}^{(1)}$ and $\mathcal{N}^{(2)}$,  
with cardinalities $N^{(1)}$ and $N^{(2)}$.
We consider the simplest case in which
each agent can be in two possible status. We can therefore
formulate the model through Ising variables $\sigma_i=\pm 1$, with $i\in \mathcal{N}^{(1)}\cup \mathcal{N}^{(2)}$. 
According to the sign of the coupling, friendly or unfriendly, 
each agent $i$ tends to follow or anti-follow its neighbors, by minimizing or maximizing the term
$\sigma_i\sum'\sigma_j$, where $\sum'$ runs over the set of neighbors of $i$.
When two communities are involved it is necessary to distinguish between intra- and inter-couplings.
We hence define the $2\times 2$ matrix $J^{(l,m)}$: 
$J^{(1,1)}$ and $J^{(2,2)}$ are the intra-couplings, and $J^{(1,2)}=J^{(2,1)}$ the inter-coupling. 
Furthermore, for the most general formulation
we should introduce also the $2\times 2$ matrix $\Gamma^{(l,m)}$
defined as the set of coupled spins $(i,j)$ within the same community (intra), or between the two communities (inter).
Finally, we introduce a global factor $\beta$ that rescales all the couplings in $\beta J^{(l,m)}$.

From the equilibrium statistical mechanics viewpoint, above we have defined
a Ising model on communities \cite{Comm}. 
However, we will make use of a discrete-time ``Glauber dynamics'' that,
when unfriendly couplings are involved, has little to share with classical results of statistical mechanics.
In fact, the continuous-time Glauber dynamics \cite{Glauber,Redener} and its discrete-time version may be totally different.
Such a difference was already raised in \cite{DPotts} in the context of the Potts model for one community.
In that case we showed that an unfriendly coupling and a discrete-time dynamics result, asymptotically, in permanent oscillations of period 2.
{However, in any 1-community model governed by quadratic interactions (Ising, Potts, Vector model etc ...), only period 2 oscillations may exist.
In this Letter}, we go beyond one single community and we show that oscillations of any period, or even totally aperiodic, can arise
as a result of the quadratic interaction between the two communities. 
More precisely, for such a scenario, in the case of quadratic interactions
three conditions must be realized:
\textit{i)} the presence of unfriendly couplings, \textit{ii)} the discrete-time nature of the dynamics,
and \textit{iii)} the existence of at least two interacting communities which rearrange their configurations
at alternate (discrete) times. We will see that \textit{iii)} leads to effective non-linear intra-couplings and then,
via \textit{i)} and \textit{ii)}, to chaos. 
Chaos is in fact a rather common feature in non-linear discrete-time dynamical systems \cite{Logistic},
and in statistical mechanics 
was observed in systems with competitive nearest- and next-nearest couplings \cite{Salinas}, as well as
in spin glass models \cite{Alava,Berker}. By contrast,
we notice that not much emphasis has been given to chaos 
in modeling social dynamics \cite{SantoS}, perhaps due to the fact that, to the best of our knowledge, 
the social dynamics considered so far did not account for all the conditions \textit{i)}-\textit{iii)}.
{Note that 1-community models may develop chaos, but explicit non linear interactions (\textit{i.e.}, non quadratic, in potential energy terms)
must be imposed; as done in \cite{Bagnoli}, where each agent can act both friendly or unfriendly depending on its neighbors. Our work instead shows that chaos turns out to be a much
more universal feature.}
We point out also that, while \textit{ii)}
is certainly absent in the classical realm of physics, where the time is continuous,
\textit{i)}-\textit{iii)} are typical in many human-based systems, and also in eco-systems, where each agent takes action at certain
discrete random times as a result of the changes of its neighbors in the same and in the other community, and the action 
may be competitive/unfriendly.
As an example, consider the US and Asian stock markets in which the spin state provides the will to sell or buy 
derivatives of a specific good at two specific prices. 
In each market the tendency of each agent is influenced by the other agents in the same market.
But also, each market influences the other market. On the other hand actions in each market take
place at discrete times. 
Furthermore, since the US and the Asian markets are located in different time zones,
their respective rearrangements
take place at different times. 
\begin{figure}[htb]
  \begin{center}
    \begin{tabular}{l}
      {\includegraphics[width=8.1cm]{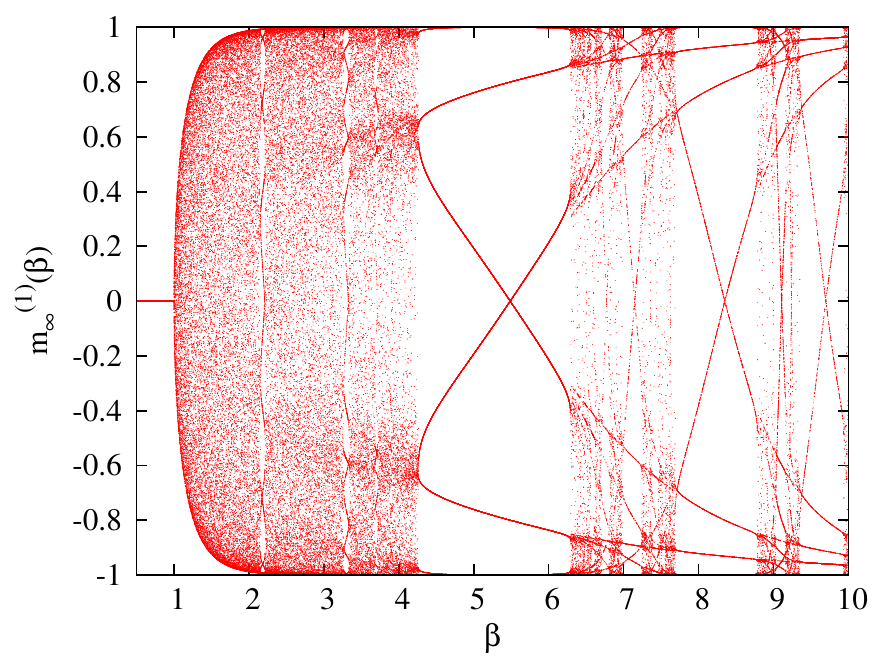}} \\
      {\includegraphics[width=8.1cm]{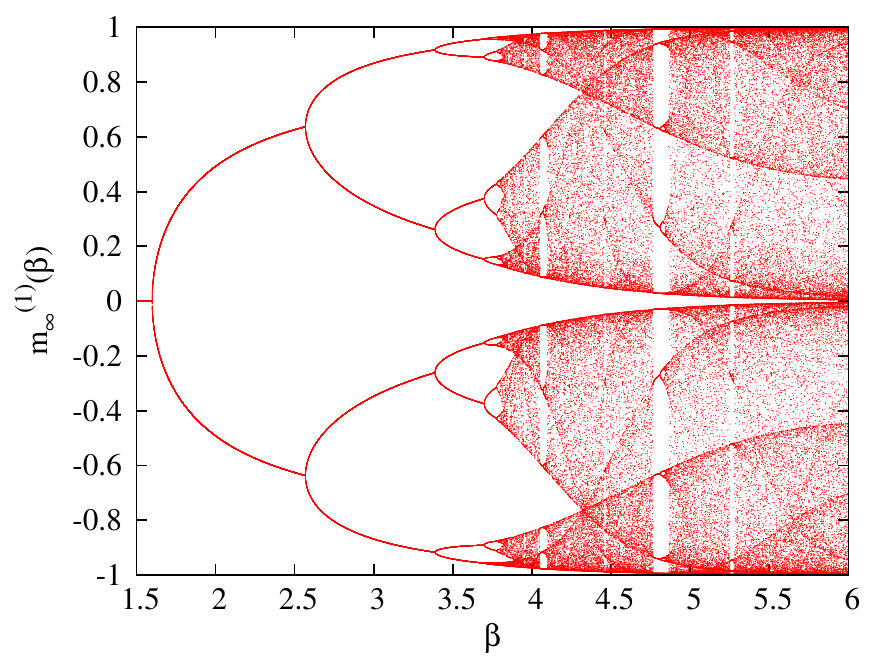}} 
    \end{tabular}
    \caption{(Color online) Plot of $m_\infty^{(1)}(\beta)$, \textit{i.e.}, 
      $m^{(1)}(t_l)$ (Eqs. (\ref{mastm1})-(\ref{mastm2}); here with $\alpha=1$) as a function of $\beta$, with $t_l$ high enough to remove
      temporary transients. In this case $t_l=10^4$.
      Each value of $\beta$ is in correspondence with a random initial condition for a total of $10^5$ samples.  
      The first bifurcation point coincides with the critical $\beta$ solution of Eq.~(\ref{crit}).
      Top: 
      $\tilde{J}^{(1,1)}=\tilde{J}^{(2,2)}=-1$, $\tilde{J}^{(1,2)}=\tilde{J}^{(2,1)}=0.8$.
      Bottom: 
      a system with 
      $\tilde{J}^{(1,1)}=-1$, $\tilde{J}^{(2,2)}=0.01$, $\tilde{J}^{(1,2)}=\tilde{J}^{(2,1)}=1$.
  \label{fig1}
    }
  \end{center}
\end{figure}

Such a scenario can be formalized as follows.
Let us indicate the discrete time by $t$. It is not restrictive to suppose that $t$ belongs to the set of natural numbers $\NN$.
Let $(\PN,\QN)$ be any periodic partition of $\NN$, \textit{i.e.}, $\PN\cup \QN=\NN$, $\PN\cap\QN=\emptyset$, and 
$\PN$ and $\QN$ contain both infinite elements of $\NN$. For instance, $\PN$ and $\QN$ can be the set of even and odd numbers, respectively,
and we will assume this choice in all the next examples.
We now introduce the following local transition rate probabilities for the spin with state $\sigma_i$ to jump to the state $\sigma_i'$ (see Eqs. (\ref{rate})).
\begin{widetext}
\begin{eqnarray}
\label{rate}
w(\sigma_i\to\sigma_i';t)=   
\left\{ 
\begin{array}{l}
\delta_{\sigma_i,\sigma_i'} 
\quad i\in\mathcal{N}^{(1)}, \quad \mathrm{for~}t\in \QN, \\ \\
\frac{1+\sigma_i'\tanh\left(\beta J^{(1,1)}\sum_{j:(i,j)\in \Gamma^{(1,1)}}\sigma_j+
\beta J^{(1,2)}\sum_{j:(i,j)\in \Gamma^{(1,2)}}\sigma_j\right)}{2}, 
\quad \\ \mathrm{for~} i\in\mathcal{N}^{(1)}, \quad t\in \PN,\\ \\
\delta_{\sigma_i,\sigma_i'}
\quad i\in\mathcal{N}^{(2)}, \quad \mathrm{for~} t\in \PN,\\ \\
\frac{1+\sigma_i'\tanh\left(\beta J^{(2,2)}\sum_{j:(i,j)\in \Gamma^{(2,2)}}\sigma_j+
\beta J^{(2,1)}\sum_{j:(i,j)\in \Gamma^{(2,1)}}\sigma_j\right)}{2}, 
\quad \\ \mathrm{for~} i\in\mathcal{N}^{(2)}, \quad t\in \QN,
\end{array}
\right. 
\end{eqnarray} 
\end{widetext}

In principle other choices are possible.
From the viewpoint of modeling, Eqs. (\ref{rate}) are justified as 
they make each spin to follow the majority of its intra- and inter-neighbors and,
thanks to the presence of the functions $\tanh(\cdot)$, the rates are non negative
and normalized at any time $\sum_{\sigma'}w(\sigma\to\sigma';t)=1/$(Time Unit).
From a deeper viewpoint, Eqs. (\ref{rate})
are based on the fact that, as we will see soon, in the case of positive couplings 
they lead to Boltzmann equilibrium governed by only quadratic interactions.
More precisely, in the case of positive couplings
the form (\ref{rate}) guarantees that at equilibrium the system satisfies 
the principle of detailed balance and the principle of
maximal entropy for any quadratic interactions.

We formalize the discrete-time probabilistic dynamics induced by Eqs. (\ref{rate}) as follows. 
Let $N=N^{(1)}+N^{(2)}$. Let us introduce the 
spin vector $\bm{\sigma}=(\sigma_1,\ldots,\sigma_N)$, and the 
associated probability vector $p(\bm{\sigma};t)$,
\textit{i.e.}, the probability that the system
is in the configuration $\bm{\sigma}$ at time $t\in \NN$.
The master equation reads
\begin{eqnarray}
\label{Master1}
&& \frac{p(\bm{\sigma};t+1)-p(\bm{\sigma};t)}{\alpha}=
-\sum_{\bm{\sigma}^{'}}p(\bm{\sigma};t)W(\bm{\sigma}\to\bm{\sigma}^{'}) ~~~~ \nonumber \\ && 
+\sum_{\bm{\sigma}^{'}}p(\bm{\sigma}^{'};t)W(\bm{\sigma}^{'}\to\bm{\sigma}),
\end{eqnarray}
where we have introduced the global transition rates
\begin{eqnarray}
\label{W}
W(\bm{\sigma}\to\bm{\sigma}^{'})=\prod_{i\in \mathcal{N}^{(1)}\cup \mathcal{N}^{(2)}} w(\sigma_i\to\sigma_i'),
\end{eqnarray}
and where $\alpha/2>0$ may be interpreted as the rate at which,
due to the interaction with an environment,
a free spin ($J^{(l,m)}=0$) makes transitions from either state to
the other. As we have proved in \cite{DPotts}, it is necessary
to impose the bound $\alpha\leq 1$ for $p(\bm{\sigma};t)$ to 
be a probability at any time $t$.
By using Eqs. (\ref{rate})-(\ref{W}) it is easy to check that the stationary solutions
$p(\bm{\sigma})$ of Eq. (\ref{Master1}) are given by the Boltzmann distribution 
$p(\bm{\sigma})\propto \exp[-\beta H (\bm{\sigma})]$, where 
\begin{eqnarray}
\label{H}
H= && - J^{(1,1)}\sum_{(i,j)\in\Gamma^{(1,1)}}\sigma_i\sigma_j-J^{(2,2)}\sum_{(i,j)\in\Gamma^{(2,2)}}\sigma_i\sigma_j\nonumber \\ &&
      - J^{(1,2)}\sum_{(i,j)\in\Gamma^{(1,2)}}\sigma_i\sigma_j.
\end{eqnarray}
It is worth to remind that the existence of a stationary solution $p(\bm{\sigma})$
does not represent a sufficient condition for equilibrium.
In fact, analogously to the case of one single community \cite{DPotts}, 
we will see that when anti-cooperative couplings are involved,
asymptotically the system can reach non-point-like attractors, if any.

Eqs. (\ref{rate})-(\ref{W}) define the microscopic dynamics from which
one can derive the macroscopic (or reduced)
dynamics, \textit{i.e.}, the dynamics for the order parameters
\begin{eqnarray}
\label{mm1}
&& m^{(1)}(t)=\sum_{\bm{\sigma}}p(\bm{\sigma};t)\frac{1}{N}\sum_{i\in\mathcal{N}^{(1)}} \sigma_i, \\
\label{mm2}
&& m^{(2)}(t)=\sum_{\bm{\sigma}}p(\bm{\sigma};t)\frac{1}{N}\sum_{i\in\mathcal{N}^{(2)}} \sigma_i.
\end{eqnarray}

\textit{The mean-field limit.}
Eqs. (\ref{rate})-(\ref{W}) can lead to
very interesting patterns and phase transitions.
For finite dimensional systems non trivial simulations would be necessary
to investigate the details of both the microscopic and macroscopic dynamics. 
In this Letter we want to focus on the mean-field limit defined by the settings
$|\Gamma^{(1,1)}|=\binom{N^{(1)}}{2}$, $|\Gamma^{(2,2)}|=\binom{N^{(2)}}{2}$,
$|\Gamma^{(1,2)}|=N^{(1)}N^{(2)}$, and the replacements
$J^{(1,1)}\to J^{(1,1)}/N^{(1)}$, $J^{(2,2)}\to J^{(2,2)}/N^{(2)}$, $J^{(1,2)}\to J^{(1,2)}(N^{(1)}+N^{(2)})/(2N^{(1)}N^{(2)})$.
If we parametrize the size of the two communities as 
\begin{eqnarray}
\label{aa}
N^{(1)}=N\rho^{(1)}, \quad N^{(2)}=N\rho^{(2)}, \quad \rho^{(1)}+\rho^{(2)}=1,
\end{eqnarray}  
for $N\to\infty$ in the rhs of Eqs. (\ref{rate}) we can apply the strong law of large numbers and the rates
simplify as
\begin{eqnarray}
\label{ratemf}
&& w(\sigma_i\to\sigma_i';t)= \\ &&
\left\{
\begin{array}{lrr}
\delta_{\sigma_i,\sigma_i'}, \quad \quad
~~~~~~~~~~~~~~~~~~~~~~~~~~~~~~~~~~ i\in\mathcal{N}^{(1)},~ t\in \QN, \nonumber \\ 
\frac{1+\sigma_i'\tanh\left(\beta \tilde{J}^{(1,1)}m^{(1)}(t)+\beta \tilde{J}^{(1,2)}m^{(2)}(t)\right)}{2}, 
~i\in\mathcal{N}^{(1)},~ t\in \PN,\\ \\
\delta_{\sigma_i,\sigma_i'}, \quad \quad
~~~~~~~~~~~~~~~~~~~~~~~~~~~~~~~~~~ i\in\mathcal{N}^{(2)}, ~ t\in \PN,\\ 
\frac{1+\sigma_i'\tanh\left(\beta \tilde{J}^{(2,2)}m^{(2)}(t)+\beta \tilde{J}^{(2,1)}m^{(1)}(t)\right)}{2},  
~ i\in\mathcal{N}^{(2)},~t\in \QN,
\end{array}
\right. 
\end{eqnarray}  
where the matrix $\bm{\tilde{J}}$ is given by
\begin{eqnarray}
\label{Jtilde}
\bm{\tilde{J}}=\left(
\begin{array}{ll}
 J^{(1,1)} & \frac{J^{(1,2)}}{2\rho^{(1)}} \\
\frac{J^{(2,1)}}{2\rho^{(2)}} & J^{(2,2)}
\end{array}
\right).
\end{eqnarray}  
By plugging Eqs. (\ref{ratemf}) in (\ref{Master1}) we get 
the following deterministic evolution Eqs. for the order parameters (\ref{mm1})-(\ref{mm2})
\begin{eqnarray}
\label{mastm1}
&&\frac{m^{(1)}(t+1)-m^{(1)}(t)}{\alpha}=\\ &&
\left\{
\begin{array}{lll}
0,  \quad t\in \QN, \nonumber \\ \\
\tanh\left(\beta \tilde{J}^{(1,1)}m^{(1)}(t)+\beta \tilde{J}^{(1,2)}m^{(2)}(t)\right)-m^{(1)}(t),\\ ~t\in \PN,
\end{array}
\right.
\end{eqnarray}  
\begin{eqnarray}
\label{mastm2}
&&\frac{m^{(2)}(t+1)-m^{(2)}(t)}{\alpha}=\\ &&
\left\{
\begin{array}{lll}
0,  \quad t\in \PN,\nonumber \\ \\
\tanh\left(\beta \tilde{J}^{(2,2)}m^{(2)}(t)+\beta \tilde{J}^{(2,1)}m^{(1)}(t)\right)-m^{(2)}(t),\\~t\in \QN.
\end{array}
\right.
\end{eqnarray}  

\textit{Friendly couplings.}
As mentioned before, when the couplings are positive,
there is little difference between the present discrete-time dynamics
and the continuous Glauber dynamics, and both asymptotically reach 
equilibrium according to the Boltzmann distribution  
$p(\bm{\sigma})\propto \exp[-\beta H_{\mathrm{mf}} (\bm{\sigma})]$, where 
$H_{\mathrm{mf}} (\bm{\sigma})]$ is the mean-field analogous of Eq. (\ref{H}),
and $m^{(1)}(t)$ and $m^{(2)}(t)$ tend, for $t\to\infty$, to the stationary solutions of Eqs. (\ref{mastm1})-(\ref{mastm2}), \textit{i.e.},
\begin{eqnarray}
\label{mstat}
\left\{
\begin{array}{lll}
m^{(1)}=\tanh\left(\beta \tilde{J}^{(1,1)}m^{(1)}+\beta \tilde{J}^{(1,2)}m^{(2)}\right),\\
m^{(2)}=\tanh\left(\beta \tilde{J}^{(2,1)}m^{(1)}+\beta \tilde{J}^{(2,2)}m^{(2)}\right).
\end{array}
\right.
\end{eqnarray}  

Eqs. (\ref{mstat}) represent a particular case of the general result derived in \cite{Comm}
valid for $n$ interacting communities at equilibrium.
In particular, one can check that Eqs. (\ref{mstat}) give rise to  
second order phase transitions whose critical surface is determined by the condition
\begin{eqnarray}
\label{crit}
\det\left(\bm{1}-\beta\bm{\tilde{J}}\right)=0.
\end{eqnarray}  
In general, $(m^{(1)},m^{(2)})=(0,0)$ is stable when the eigenvalues of $\beta \bm{\tilde{J}}$ 
are inside the interval $(-1,1)$ (paramagnetic phase, or no consensus),
otherwise the system reaches a spontaneous magnetization $(m^{(1)},m^{(2)})\neq (0,0)$ (frozen phase, or consensus). 

\begin{figure}[thb]
\includegraphics[width=7.1cm]{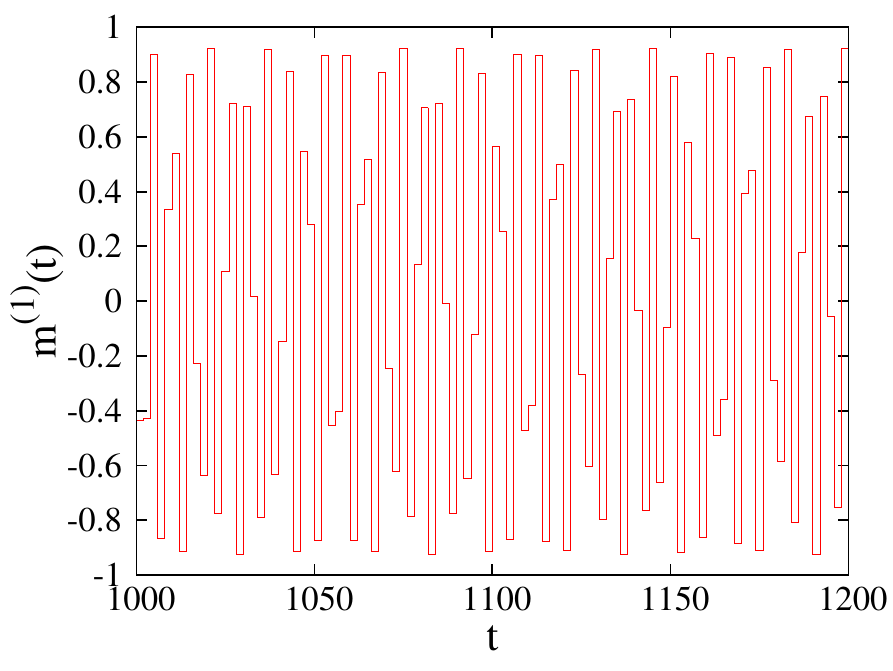}
\includegraphics[width=7.1cm]{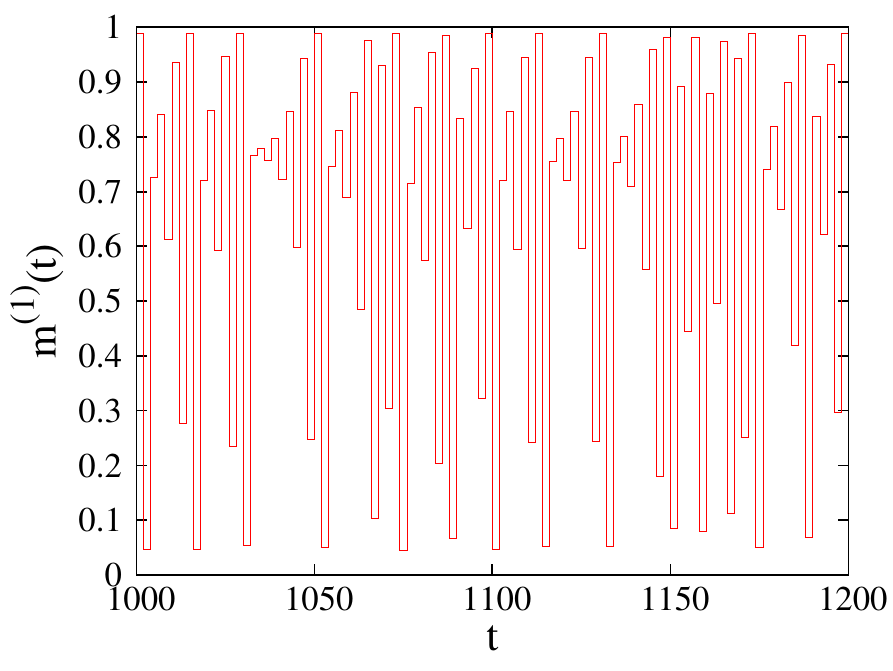}
\caption{(Color online) 
Top: A trajectory of $m^{(1)}(t)$ from the case of Top Panel of Fig. \ref{fig1} at $\beta=1.5$.
Bottom: A trajectory of $m^{(1)}(t)$ from the case of Bottom Panel of Fig. \ref{fig1} at $\beta=4.5$. The two trajectories reflect the two different scenarios: marginal chaos characterized by aperiodic trajectories and exponential dependence on the model parameters for the former, and true chaos characterized by strong intermittency and exponential dependence on the initial conditions for the latter. The two scenarios are in correspondence with zero and positive maximum Lyapunov exponent, respectively.
\label{fig2}
}
\end{figure}

\textit{Competitive interactions; emergence of chaos.}
In general, when one or more couplings are negative the discrete-time dynamics
has no resemblance with the continuous-time dynamics, and none of the arguments
valid to analyze the latter should be used for the former. As we have discussed in the Introduction,
when the conditions \textit{i)}-\textit{iii)} are verified,
oscillations of any period can take place. 
Depending on the value of $\bm{\tilde{J}}$ and on $\beta$, the system can reach many different
regimes with fixed points, limit cycles, or strange attractors.
Given $\bm{\tilde{J}}$, the best way to visualize such a complex scenario is to let the system
to evolve toward high enough values of $t=t_l$ in order to remove
temporary transients and repeat the numerical experiment for several values of $\beta$, each 
$\beta$ being associated to a random initial condition.
Hence, we plot $(m^{(1)}(t_l),m^{(2)}(t_l))$ as functions of $\beta$, and we indicate
these functions as $(m_\infty^{(1)}(\beta),m_\infty^{(2)}(\beta))$. 
In our examples $t_l \geq 10^3$ turns out to be enough high. 
For the functions $(m_\infty^{(1)}(\beta),m_\infty^{(2)}(\beta))$ the variable $\beta$ plays the role of a time-scaling:
the higher $\beta$, the higher $t_l$. In general, the functions $(m_\infty^{(1)}(\beta),m_\infty^{(2)}(\beta))$ look multi-valued
functions due to the existence of bifurcation points $\beta^*$ where they
undergo a bifurcation. There exist two kinds of bifurcations: 
doubling period and phase transition.
In the former case, if for $\beta<\beta^*$ the functions 
$(m^{(1)}(t),m^{(2)}(t))$ follow periodic oscillations of period $T$, at $\beta^*+\delta\beta$
with $\delta\beta>0$, they will follow periodic oscillations of period $2T$.
In the other case, the bifurcation corresponds to a broken symmetry and, depending on the
initial condition, for $\beta=\beta^*+\delta\beta$ a single trajectory will follow either the upper or the lower branch.
In particular, the first bifurcation 
coincides with the critical point $\beta_c$ of the equilibrium Ising model determined by Eq. (\ref{crit}). 
Given a bifurcation point it is always possible to distinguish whether it corresponds to a doubling period or to a 
phase transition. It is in fact enough to plot $(m_\infty^{(1)}(\beta),m_\infty^{(2)}(\beta))$ by choosing
the same initial condition for each $\beta$: the bifurcations in such a plot will correspond only
to doubling period. 
However, it is clear that in either case, doubling period or phase transition, the presence of bifurcations
increases the chance to develop chaos. Therefore, for our aims here,
it is more interesting, as well as highly more efficient, to show the plots 
that include all the bifurcation points. 
Figs. \ref{fig1} (Top and Bottom) show two different scenarios and Figs. \ref{fig2} (Top and Bottom) two corresponding sample trajectories.
Below we provide a concise description of these two scenarios. Detailed calculations will be reported elsewhere \cite{Elsewhere}. 

The first scenario takes place when each community is anti-cooperative but the two are mutually cooperative. 
Rigorously speaking, in this case the system is not chaotic, but rather ``marginally chaotic''. 
In fact, the maximal Lyapunov exponent \cite{StrogatzB} $\lambda_1$ is zero in correspondence of the hazy regions
of Top of Fig. \ref{fig1}, while elsewhere is negative \cite{Elsewhere}. The fact that $\lambda_1=0$ means that the trajectories are not chaotic,  
yet \textit{i)} they are aperiodic, and \textit{ii)} $m^{(1)}$ and $m^{(2)}$ are dense in an interval $(a,b)\subset [-1,1]$,
where $a$ and $b$ depend on $\beta$ (in the example of Top of Fig. \ref{fig1}, for $2\leq \beta\leq 4$ we have $a=-1$ and $b=1$).
In this case scenario, unlike a true chaotic behavior, there is no high sensitiveness to the initial conditions, 
yet the system is highly sensitive to small changes of $\beta$ \cite{Elsewhere}.

\begin{figure}[thb]
\includegraphics[width=7.1cm]{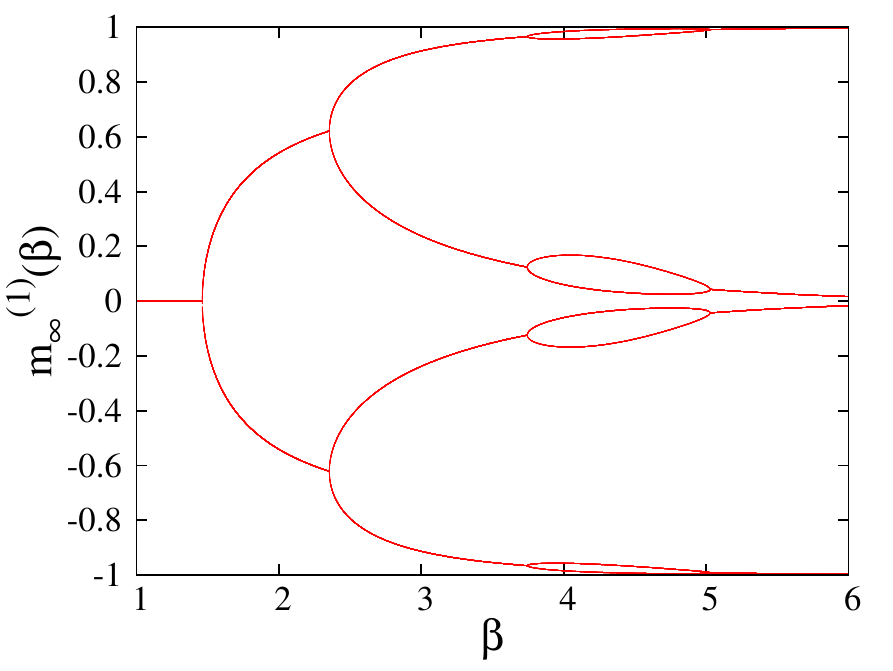}
\caption{(Color online) Plot of $m_\infty^{(1)}(\beta)$ 
for a system with $\alpha=1$,
$\tilde{J}^{(1,1)}=-1$, $\tilde{J}^{(2,2)}=0.09$, $\tilde{J}^{(1,2)}=\tilde{J}^{(2,1)}=1$.
\label{fig3}
}
\end{figure}

The second scenario takes place when there is
an anti-cooperation in one community and a small or null cooperation in the other. It is characterized by bifurcation cascades
and a few windows of stability, and the whole structure turns out to be similar to the chaos of a symmetrized version 
of the logistic map \cite{Logistic}. In fact, in this case we have $\lambda_1>0$ in correspondence of the hazy regions of Bottom of Fig. \ref{fig1} \cite{Elsewhere},
which means that, there, the trajectories are chaotic, characterized by a strong intermittency \cite{Vulpiani}, as seen in Bottom of Fig. \ref{fig2}.
We can understand the mechanism leading to chaos by exploiting the limit $\tilde{J}^{(2,2)}\to 0$. In this limit Eqs. (\ref{mastm1}) and (\ref{mastm2})
decouple and, for example, for $\alpha=1$ we are left with
\begin{eqnarray}
\label{dec}
&& m^{(1)}(t+2)=\\
&& \tanh\left(\beta \tilde{J}^{(1,1)}m^{(1)}(t)+\beta \tilde{J}^{(1,2)}\tanh\left(\beta \tilde{J}^{(2,1)}m^{(1)}(t)\right)\right). \nonumber
\end{eqnarray}  
Eq. (\ref{dec}) shows that $m^{(1)}(t)$ "feels'' an effective non linear coupling due to the presence of the second community,
even though there is no coupling within it. 
However, even small values of $\tilde{J}^{(2,2)}$ can remove the chaos, see Fig. \ref{fig3}.

\textit{Conclusions - considerations on stock markets.} 
{In recent years, has become clear that the presence of 2 or more communities greatly affects the behavior of a model.
In particular, studies at equilibrium have shown that two or more communities lead to the existence of several unstable and metastable states
not seen in the 1-community case \cite{Comm,Suchecki2}. Such thermodynamically negligible states, however, can play a dramatic role at the dynamical level.}

In this Letter, we have introduced a simple model of two interacting communities 
subject to stochastic discrete-time updates and governed by only quadratic interactions.
When the intra- and inter-couplings are cooperative,
the two communities reach asymptotically an equilibrium state.
However, when the intra- or inter-couplings are anti-cooperative, 
the system may remain in perpetual oscillations and,
when the coupling values belong to certain intervals, 
two possible scenarios arise, characterized either by erratic aperiodic trajectories and high sensitiveness to small changes of the couplings,
or by chaotic trajectories and bifurcation cascades.
Quite interestingly, we find that even a moderate consensus in one single community
can remove the chaos. 
Despite the related Hamiltonian (\ref{H}) contains only bilinear interactions, the system can develop 
a marginally- or a fully-chaotic behavior. 
This result implies: when at least two communities are involved, chaos is more likely to occur. 
The plain model considered here 
provides a minimal mechanism for chaos based on the presence of interacting communities 
and calls for more realistic models,
especially in view of applications to social dynamics \cite{SantoS}, economics, and ecosystems, 
where the conditions \textit{i)}-\textit{iii)} discussed in the Introduction
turn out to be typical. 
We stress that the dynamics that we have used is based on a majority rule that does not involve explicit non-linear terms;
it is the presence of at least two different communities that produces effective non-linear terms
responsible for erratic trajectories and a chaotic behavior.
An open interesting question is whether this chaotic behavior persists also beyond the mean-field limit.  
\begin{figure*}[thb]
  \includegraphics[width=\textwidth,height=6.2cm]{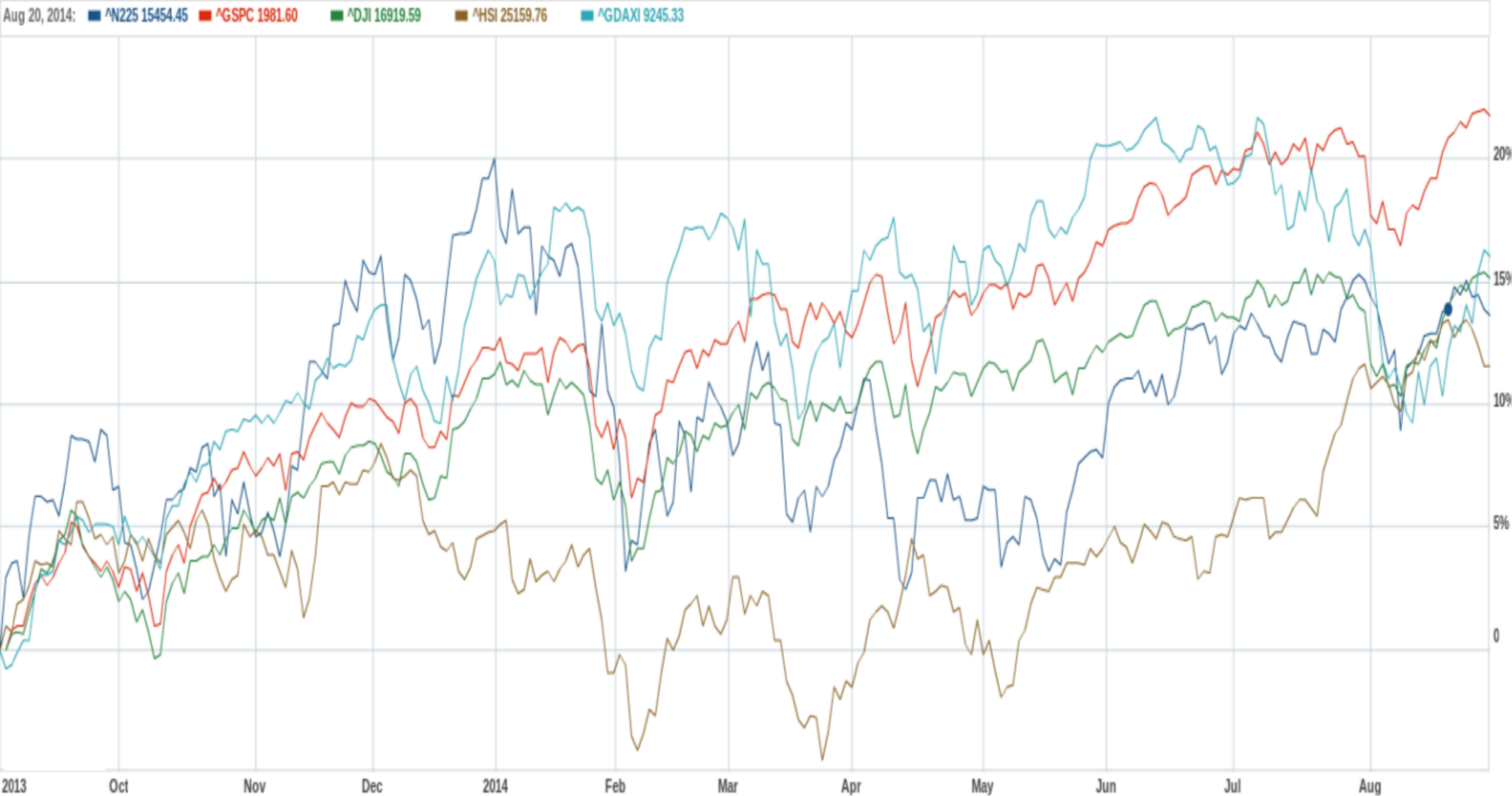}
\caption{(Color online) Comparison over 1 year of 5 industrial (mainly) stock indices: Japan Nikkei-225 (Blue), U.S. stocks S\&P-500 (Red) 
and Dow Jones (Green), Hong-Kong HSI (Brown),
and German Dax (Cyan) (source CSI). \label{fig4}}
\end{figure*}
As mentioned in the Introduction, our analysis finds a real-world counterpart 
within interacting stock markets, where the observed chaotic behavior can be explained in terms of 
agents each following (or anti-following) the behavior of the majority of its intra- or inter-neighbors, the latter 
operating on different stock markets with different time-zones.  
Fig. \ref{fig4} shows the behavior of the Japanese, Hong-Kong, German and two US stock markets.
The Asian and US stock markets operate in two non overlapping working times, whereas the German market has time overlaps
with both the Asian and the US markets. 
We observe that markets with no overlapping time are not able to reach any mutual equilibrium (positive correlated behavior),
while a certain mutual equilibrium seems possible within markets having a full (S\&P-500 and Dow Jones) or almost full (Nikkei 225 and Hang-Seng)
overlapping time.
According to our model, the lack of mutual equilibrium can be
seen as due to the presence of negative couplings (for example currency changes can trigger
such anticorrelations). 
Of course the correlated behavior might be partially due to external factors. In fact, geo-political factors
are responsible for the larger fluctuations which could be incorporated in the model via
suitable stochastic external fields. Other interesting issues could be at least qualitatively addressed by the model
and will be the subject of future works.

\begin{acknowledgments}
Work supported 
by CNPq Grant PDS 150934/2013-0.
W. F. also acknowledges CNPq, CAPES and FAPESC.
We thank G. Baxter and C. Presilla for useful discussions.
\end{acknowledgments}


\end{document}